%% file: main.tex
\def\year{2021}\relax
\documentclass[letterpaper]{article} 
\usepackage{aaai21}  
\usepackage{times}  
\usepackage{helvet} 
\usepackage{courier}  
\usepackage[hyphens]{url}  
\usepackage{graphicx} 
\usepackage{xcolor}
\usepackage{xspace}

\usepackage{natbib}  
\usepackage{caption} 
\usepackage{xspace}
\usepackage{dirtytalk}
\usepackage{comment}
\usepackage{enumitem}

\newcommand{\system}{\textsc{EchoPal}\xspace}

\newcommand{\worker}{worker\xspace}
\newcommand{\workers}{workers\xspace}

\usepackage{booktabs,siunitx}

\newenvironment{choruschat}
    {
        \begin{enumerate}[leftmargin=3pc,style=nextline,align=right,itemsep=0mm]
        \small
        \sffamily
        \vspace{.3pc}
    }
    {
        \vspace{.3pc}
        \end{enumerate}
    }
\newcommand{\echo}[1]{\item [\textbf{Echo:}]\xspace \textbf{#1}} 
\newcommand{\user}[1]{\item [User:]\xspace #1} 
\newcommand{\voiceover}[1]{\item []\xspace [\textit{#1}]} 

\newcommand{\kenneth}[1]{}

\title{Too Slow to Be Useful?\\On Incorporating Humans in the Loop of Smart Speakers}

\author {
    Shih-Hong Huang,~
    Chieh-Yang Huang,~
    Yuxin Deng,\\
    Hua Shen,~
    Szu-Chi Kuan,~
    Ting-Hao `Kenneth' Huang\\
}

\affiliations {
    College of Information Sciences and Technology, The Pennsylvania State University\\
201 Old Main, University Park, PA 16802, USA\\
    \texttt{\{szh277,chiehyang,ybd5063,huashen218,sbk5672,txh710\}@psu.edu}
}

\begin{document}
\maketitle
\thispagestyle{plain}
\pagestyle{plain}

\textit{\textcolor{blue}{This is the extended technical report of the ``Too Slow to Be Useful? On Incorporating Humans in the Loop of Smart Speakers'' paper~\cite{huang2022too}, accepted by the Works-in-Progress and Demonstration track of the 10th AAAI Conference on Human Computation and Crowdsourcing (HCOMP 2022 WiP/Demo).}}
\bigskip

\begin{abstract}
\input{abstract.tex}

\end{abstract}

\section{Introduction}

\input{introduction.tex}

\section{Related Works}

\input{related-work.tex}

\section{\system System} \label{system}
\input{03_system}

\section{User Study}
\input{04_user_study}

\section{Discussion}
\input{05_discussion}

\section{What's Next?}
\input{next-step.tex}

\section{Conclusion and Future Work}
\input{06_conclusion}

\bibliography{echopal}
\end{document}

%% file: abstract.tex
Real-time crowd-powered systems, such as Chorus/Evorus, VizWiz, and Apparition, have shown how incorporating humans into automated systems could supplement where the automatic solutions fall short.
\kenneth{How does this work?}However, one unspoken bottleneck of applying such architectures to more scenarios is the longer latency of including humans in the loop of automated systems.
For the applications that have hard constraints in turnaround times, human-operated components' longer latency and large speed variation seem to be apparent deal breakers.
This paper explicates and quantifies these limitations by using an human-powered text-based backend to hold conversations with users through a voice-only smart speaker.
Smart speakers must respond to users' requests within seconds, so the workers behind the scenes only have a few seconds to compose answers.
We measured the end-to-end system latency and the conversation quality with eight pairs of participants, showing the challenges and superiority of such systems.

%% file: introduction.tex
Real-time crowd-powered systems have achieved success in reducing the gaps between human agents and automated solutions.
For example, 
VizWiz utilized crowd workers to answer visual questions quickly for blind people~\cite{bigham2010vizwiz};
Scribe asked non-experts to caption speech for deaf and hard of hearing audiences~\cite{lasecki2012real};
Zensors~\cite{laput2015zensors} and Zensors++~\cite{10.1145/3264921} to monitor running video feeds;
Chorus~\cite{lasecki2013chorus,huang2016there} and Evorus~\cite{huang2018evorus} used the crowd to hold sophisticated long conversations with users in the wild.
However, despite much effort to speeding up such systems -- for example, 
shortening the worker recruiting time~\cite{bigham2010vizwiz,bernstein2011crowds},
constraining the task completion time harshly~\cite{huang2017real},
getting help from automated components~\cite{huang2018evorus},
or trying to predict the near future so that the crowd can start working beforehand~\cite{lundgard2018bolt} -- humans are, in many cases, still slower than computers.
Many existing automated systems and their infrastructures were built with the assumption that all the internal components, when working properly, have short turnaround times.
This reality makes realizing the vision of crowd-powered systems extra challenging.

Taking modern smart speakers or voice-enabled devices for example,
Amazon's Echo devices, Google Assistant, Apple Siri, and Samsung Bixby 
respond to users' requests within around 0.77 to 3.09 seconds~\cite{10.1007/978-3-030-68452-5_39}. 
This range of turnaround time is too short for most crowd-powered systems.
The average latency of a response from the deployed version of Chorus and Evorus was longer than 30 seconds;
average response time per question for Vizwiz as 36 seconds
and the latency of Zensors++ is 120 seconds.

\begin{figure}[t]
    \centering
    \includegraphics[width=\columnwidth]{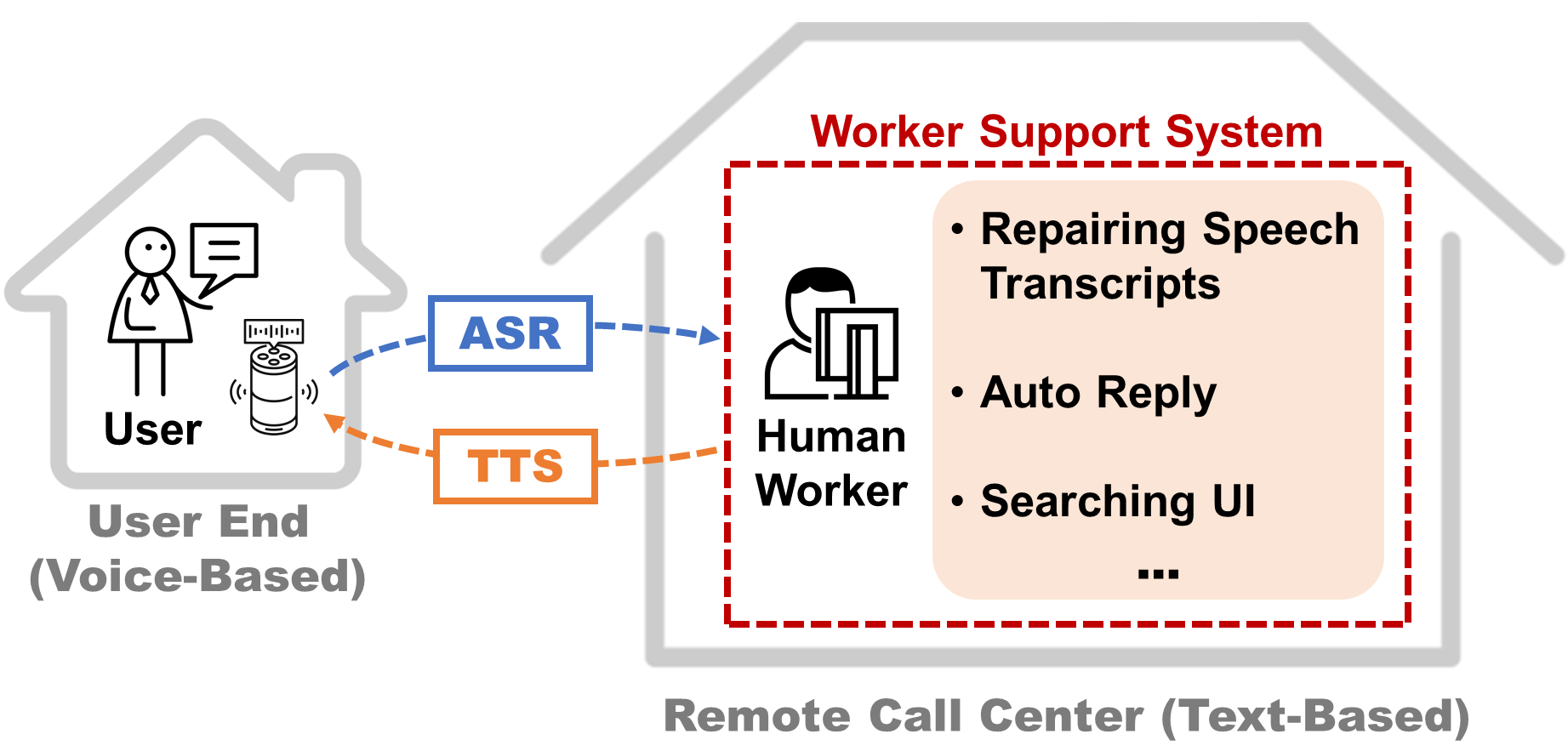}
    \caption{System overview of \system. When the user talks to \system, the speech will be transcribed and passed to the remote call center. Inside the call center, a set of automated technologies help the worker make sense of the conversation and generate responses. \system will collect the worker response and speak it out to the user.}
    \label{fig:system-overview}
\end{figure}

This paper examines and explicates the challenges of incorporating a human-in-the-loop architecture into an already-existing and widely used technological infrastructure.
In particular, we built \textbf{\system},\footnote{\system demo video: \url{https://youtu.be/iMDsX52VWGY}} a prototype system that allows a human worker to converse with the user synchronously via an Amazon's Echo device\footnote{\system was developed and tested during Apr-Dec 2019. We are aware if some changes were made on the Amazon's end after we finished our user studies.}.
Figure~\ref{fig:system-overview} overviews the system.
\system works with the full set of Amazon Alexa's infrastructure, including the official Alexa Skills Kit\footnote{Alexa Skills Kit: https://developer.amazon.com/en-US/alexa/alexa-skills-kit} and an Echo device (2nd Generation). 
To incorporate human workers in the loop and to allow free conversation, we engineered a custom back end for \system and hosted it on Amazon Web Services (AWS). 

\system inherits two real-world constraints of many commercial smart speakers that are particularly challenging for human-in-the-loop systems,
 the extremely short response time, and the purely text-based back end:

\begin{itemize}

    \item \textbf{Extremely short response time.} Alexa Skills Kit's timeout is 10 seconds.
    There are two different timeout constraints for Alexa Skills Kit, processing and ASR. First, Alexa Skills Kit enforced a maximum processing time for each turn of speech received from the users. The empirical processing time observed during the development and testing of \system was longer then 30 seconds. Second, Echo devices will start to "listen" as soon as the response of the skill has been read out. The maximum listening time is approximately 10 seconds according to our observation.  
    
    \item \textbf{Purely text-based back end.} Most smart speakers do not pass voice recordings to their back-end systems to preserve user privacy.
These devices automatically compile the user's utterances into intents, semantic slots, and transcriptions and pass only these pieces of distilled information.
Namely, in \system, the worker cannot hear the user's voice.
The worker can only rely on the transcriptions that are automatically compiled, which could be imperfect or misleading, to communicate with users.
\system thus implements several automated supports for the worker to produce high-quality responses:
the system populates possible alternative audio transcriptions to help the worker make sense out of noisy audio transcriptions, and it also suggests follow-up lines automatically.
\end{itemize}

We conducted in-lab user studies to measure the conversation quality and end-to-end system latency.
We found that, unsurprisingly, the tight time constraint is the primary challenge for human \workers,
who suggested that the search support was the most useful feature they used.
Meanwhile,
compared with automated socialbots, 
the human-human conversation still showed some
superiority in terms of perceived conversation quality, confirming the benefits of using humans in the loop.
Our work explores the possibilities and challenges of human‐in‐the‐loop smart speakers, informing the designs of future systems facing various real‐world constraints.

%% file: related-work.tex
The work is related to {\em (i)} voice user interfaces and {\em (ii)} real-time crowd-powered conversational agents.

\subsection{Challenges of Voice User Interfaces}
Voice user interfaces (VUIs) have made their way into our daily lives.
Amazon alone has sold more than 100 million Alexa-enabled devices as of early 2019~\cite{bohn_2019}, and as of September 2019, over 100,000 applications are available in Alexa Skill stores~\cite{wiggers_2019}.
Researchers have also explored potential future uses of such interfaces, such as 
in-vehicle voice assistance~\cite{Braun:2019:YSD:3290605.3300270,peissner2011can,Choi:2018:DCV:3267305.3267638} or
auditory web browsers~\cite{Sato:2011:SAV:1978942.1979353}.

Meanwhile, several challenges and limitations of such interfaces have also been revealed.
Problems caused by
poor speech recognition~\cite{Lovato:2015:STY:2771839.2771910} and natural language processing~\cite{Myers:2018:PUO:3173574.3173580} often occur, and users do not use voice-enabled devices for conversation but only for short interactions to obtain responses~\cite{Porcheron:2018:VIE:3173574.3174214}.
It is also frustrating that common strategies people used to help others understand what they are trying to express would not work for voice assistants~\cite{ackerman2000intellectual}.
Researchers also realized that ``communication breakdown'' is unavoidable between users and voice assistants, where conversation stops because parties cannot understand each other and thus the communication cannot be continued.
This is especially true for children~\cite{10.1145/3202185.3202749}, elders~\cite{10.1145/3342775.3342803} and people with hearing loss~\cite{10.1145/3311957.3359487}. 
Detecting and recovering from breakdowns in communication is essential for voice-based scenarios~\cite{doi:10.1080/10447318.2015.1093856}.


The gap between the promised talking machines and how users actually use them can be characterised as a ``social-technical gap'', {\em i.e.}, the gap between what people want to accomplish (conversing naturally) and what technology is capable of~\cite{ackerman2000intellectual}.

\subsection{Real-time Crowd-powered Conversational Agents}


Real-time crowd-powered systems have achieved success in reducing the gaps between human agents and automated solutions.
Among many prior real-time crowd-powered systems, Chorus~\cite{huang2016there,lasecki2013chorus}  Evorus~\cite{huang2018evorus} and TalkBack~\cite{10.1145/3313831.3376310} are most relevant to our work.
Launched as a Google Hangouts chatbot~\cite{huang2016there}, Chorus is a crowd-powered conversational assistant that can hold open conversations about nearly any topic~\cite{lasecki2013chorus}. 
When a user initiates a conversation, a group of crowd workers is recruited from Amazon Mechanical Turk and directed to an interface where the workers propose responses, take notes on important facts, and vote on each other's replies to identify optimal responses.
Collectively, the crowd converses with the user as a single, consistent conversational partner. 
Huang {\em et al.} later created Evorus to enable Chorus to automate itself over time~\cite{huang2018evorus}. Reitmaier {\em et al.} deployed TalkBack~\cite{10.1145/3313831.3376310}, a machine and human speech assistance, in public spaces. This study showed
that humans could provide better query responses than the
AI-powered version in a range of diverse cases, but the response time is almost 10 minutes if without previous conversation data. 

\paragraph{\system is different from Chorus and Evorus.}
\system is different from these systems for several reasons:

\begin{itemize}
    \item 
First, \textbf{smart speakers require much shorter response times.} Crowd-powered conversational agents were developed and tested as text-based chatbots, where the latency can be as long as a hundred seconds. Although TalkBack is a voice-based device and adopted a corpus based on the previous conversation to reduce the latency, it is still difficult to balance response quality and latency. On average, students respond to an instant message in 32 seconds, and people in the technical startup group respond in 105 seconds~\cite{avrahami2008waiting}.
The deployed version of Chorus reported an average turnaround time of 88.35 seconds~\cite{huang2016there}.
This level of latency is too long to be useful for voice-enable devices; \system needs to drastically shorten human-powered conversational systems' response time from nearly 1.5 minutes to a few seconds. 
\item 
Second, \textbf{the speech recognizer introduces an extra layer of noise to the system.} 
In Chorus, raw conversations are typed by the user and sent directly to human workers.
In \system, the user speaks to the Echo device, and the automatic speech recognizer transcribes the utterance to text and then sends it to the worker.
Users could be in noisy environments
or in situations where speaking clearly is difficult, making it difficult to communicate clearly with the worker. 
\item Finally, \textbf{\system is not a collaborative system.}
One core feature of Chorus is the real-time collaboration between multiple workers, while \system is designed to enable an individual worker to communicate with users via voice-enabled devices.
In this paper, we focus on the challenges of adding one person into the loop of smart speakers in a call-center fashion. 
\end{itemize}




%% file: 03_system.tex
\begin{figure*}[t]
    \centering
    \includegraphics[width=.99\textwidth]{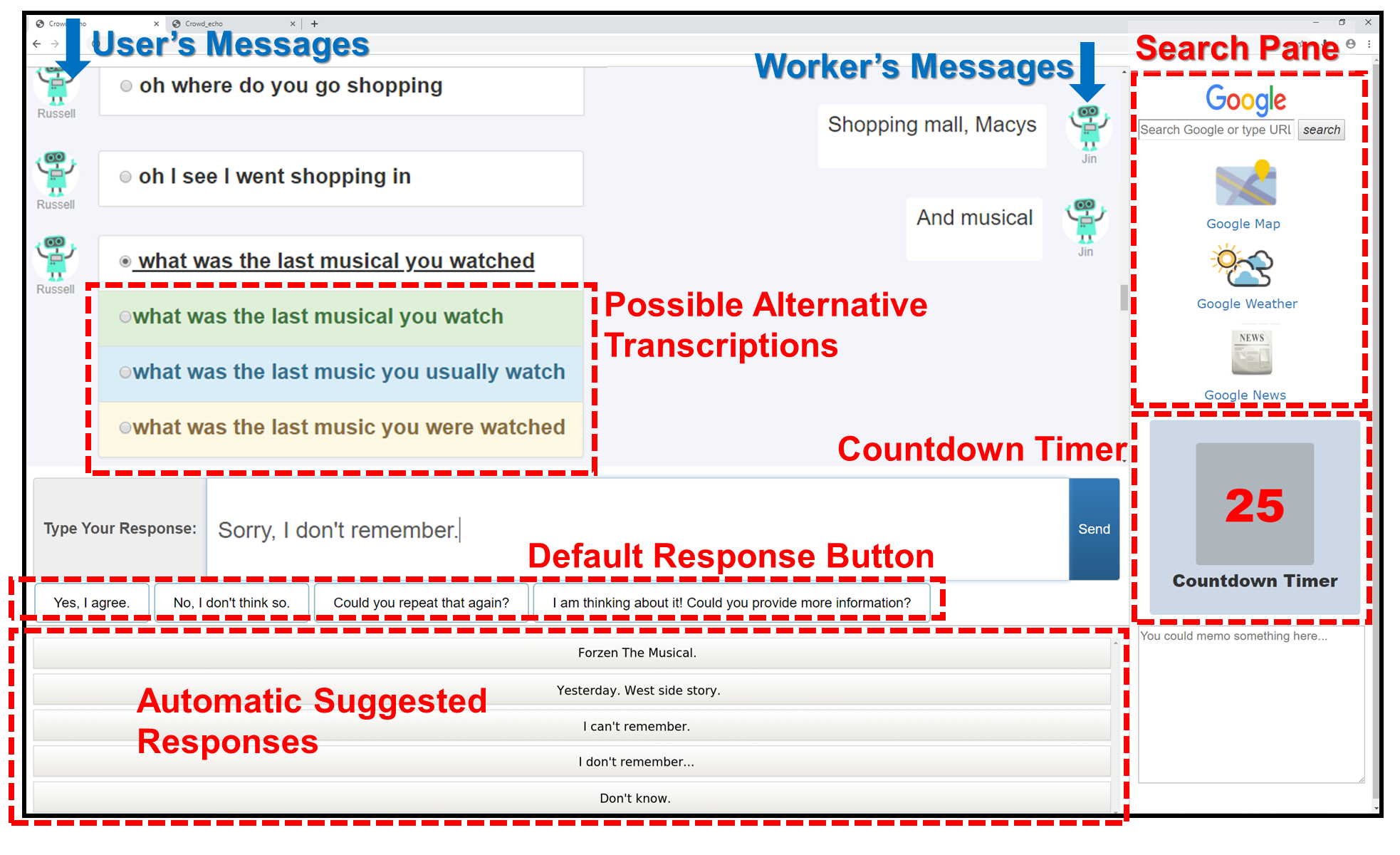}
    \caption{The worker interface of \system. When the user talks to \system, the speech will be transcribed and passed to the worker interface. Three possible alternative transcriptions and five automatic suggested responses will be presented helping the worker understand the transcription and generate response. In the search pane, \system embeds several search sites for \workers when searching is needed. \system will collect the worker response and speak it out to the user.} 
    \label{fig:worker_interface}
\end{figure*}

The workflow of \system is shown in Figure~\ref{fig:system-overview}.
When a user talks to the system, Echo records the audio and turns the speech into text through the built-in automatic speech recognition (ASR) system.
The transcribed text is then sent to the backend of \system and presented to the human worker as a message in the worker interface, where the \worker can see not only the transcribed message but also a set of possible alternative transcriptions generated by the system.
Furthermore, \system also automatically generates and presents a list of suggested responses to the \worker.
As each alternative transcription has its own suggested responses, the \worker can click on the transcription to switch between them.
Search support sites, such as Google Search or Google Weather, are shown on the right.
Equipped with these technological support features, we enforced a time constraint of 25 seconds for the \worker to produce each response, a figure empirically set by our pilot study and other related work~\cite{avrahami2008waiting}.
The \worker's response is then be sent back to the Echo device, where a built-in text-to-speech (TTS) system reads out the message to the user. In the following section, we introduce each component in detail.

\paragraph{Repurposing the Alexa Skill.}
Current configuration of Alexa Skill is not designed to take arbitrary inputs from the user.
This subsection describes our engineering procedure to enable Alexa to hold free conversation with users.

An Alexa Skill has three major components: Slots, Utterances and Intents.
Slots are lists of items within the same categories.
Utterances are sentences including one or more Slots of similar purpose.
Intents are groups of Utterances which can perform specific tasks.
For example, in a traveler assistant skill, 
the Utterances can be ``How will the weather be in New York on Monday?'' or ``Will it  be raining in Los Angeles at 2PM today?'', where the Intent is ``weather inquiry'', and the Slots are {\tt LOCATION} (city names), {\tt DATE} (month, day of the week), {\tt TIME}, and {\tt WEATHER\_CONDITION}.

A default Slot type {\tt Amazon.Literal} used to exist to take in the users speech freely without the constraint of topics. 
However, the Slot 
was officially deprecated after October 2018 without any replacement.
Since no official solutions are applicable for taking in free speeches from users, developers have been trying relentlessly for such possibilities.
Though developer communities are yet to come to a stable and guaranteed conclusion, one walk around is to build a customized {\tt catchAll} Slot~\cite{catchall}.
Comparing to other Slots of any kind, {\tt catchAll} contains a list of 156 random words instead of a specific category.
The random words include scientific names, locations, and even first names. 
The words also came in the form of English and foreign words.
No correlation can be found within the {\tt catchAll} aside from the fact that they are all non-common.
The writers suspected that {\tt catchAll} incorporates a wide variety of pronunciations which lead to its effectiveness in recognizing free speech.
Throughout the testing stage, {\tt catchAll} performed well but definitely not perfect. Misrecognition still happens from time to time.
To improve the recognition accuracy, we came up with the idea of adding common stopwords to {\tt catchAll}, resulting in a {\tt catchAll} Slot containing just under 350 words.
Though improvement can be sensed, misrecognition still occurs.

\paragraph{Entering and Exiting the \system.}
Users can trigger \system by saying \say{Alexa, open \system} and then start to talk.
When talking to Alexa, the light ring on top of Echo will light up and glimmer indicating that Alexa is listening and recording the audio.
To leave \system, saying \say{cancel} or \say{stop} will cause Alexa to close the application.

\subsection{Worker Interface}
The worker interface is shown in Figure~\ref{fig:worker_interface}.
The message pane (upper left corner) contains the history of messages.
The reply pane (bottom left corner) provides response functionality, including the typing box and suggested responses.
The search pane (right) gives links to Google services. The detail information is described as follows.

\paragraph{Message Pane.}
In the message pane, user and \worker are separated, on the left and right sides, respectively.
For an incoming message produced by the user, \system will generate three possible alternatives using the post-editing techniques. As shown in Figure~\ref{fig:worker_interface}, the first message is the original transcribed speech by the Alexa built-in ASR.
The other three messages (colored in green, blue, and yellow) are the post-edited ones.
After sending the response, only the message clicked by the \worker will be kept and treated as a history message for later reference.
To prevent the \worker from not noticing a new incoming message, \system will make a ``ding'' sound for each new message. 

\paragraph{Reply Pane.}
For each turn, the worker will have 25 seconds to reply. The countdown timer is shown right next to the typing box, colored in gray with red texts. If the worker fails to respond within 25 seconds, \system will send out the typed responses if the worker has typed some text but not yet sent anything out or will randomly choose a suggested response if the worker does not type anything. As shown in Figure~\ref{fig:worker_interface}, to help workers respond quickly, \system provides \textit{default responses} and \textit{suggested responses}:
Four default responses are intended for four different use cases:
\say{Yes, I agree.},
\say{No, I don't think so.},
\say{Could you repeat that again?}, and
\say{I am thinking about it. Could you provide more information?}.
Meanwhile, five suggested responses are generated by Cleverbot, a third-party service that automatically provides responses to any given sentence (Section~\ref{sec:response-suggestion}).
\system has a five-second lock for suggested responses to promote \workers typing their own responses.
All the suggested responses will be disabled and a gray box with ``Please type the response if you can!'' will be shown within the first five seconds after a user message is received.
To further help \workers type their responses, \system also implements a voice input system, which serves as an alternative way of typing.

\paragraph{Search Pane.}
In the search pane, \system embeds Google Search, Google Map, Google Weather, and Google News for \workers when searching is needed.
\system will make a ``dong'' sound when there is only 10 seconds left and the \worker is on the other page searching for information.
On the bottom of the search pane, \workers can type a memo for reference.

\begin{table}[t]
    \centering
    \small
    \begin{tabular}{p{3.8cm}p{3.8cm}}
    \toprule \midrule
        \multicolumn{1}{c}{\textbf{Original Sentence}} &  \multicolumn{1}{c}{\textbf{Mistranscribed Sentence}} \\ \hline
        What's your favorite \underline{scene} in the movie. & What's your favorite \underline{seen} in the movie. \\
        Not bad I have dinner \underline{with} some friends. & Not bad I have dinner \underline{was a} some friends. \\
        Have you ever tried \underline{any winter} sports before. & Have you ever tried \underline{anyone a} sports before. \\
    \midrule \bottomrule
    \end{tabular}
    \caption{Examples of the transcribing error. The original sentences are the user spoken speeches transcribed by the author. The mistranscribed sentences are transcribed by Alexa built-in ASR. We can see nonsense sentences also happen.
    }
    \label{tab:transcription-error}
\end{table}

\subsection{Repairing Speech Transcripts by Automatic Posting-Editing}
\label{sec:post-editing}
Alexa's built-in ASR does not always provide satisfactory transcription.
Table~\ref{tab:transcription-error} shows some examples where Alexa failed to transcribe the correct message.
Though techniques of improving ASR exist, these approaches often need the original audio~\cite{swarup2019improving}, which is not available to Alexa Skills.
Therefore, we introduced a novel automatic post-editing approach to \textit{repair} speech transcripts, where the algorithm finds similar sentences in the sense of sound and pronunciation.
We designed two models to get acoustically similar sentences: a retrieval-based approach and a generation-based approach.

\paragraph{Retrieval-based Approach.} 
We used The Switchboard Dialog Act Corpus~\cite{Jurafsky-etal:1997}, a telephone speech corpus with about 8,000 unique recorded sentences.
To measure acoustic similarity, we first turned all sentences into phoneme sequences using the g2pE package~\cite{g2pE2019}.
Normalized Levenshtein~\cite{levenshtein1966binary,tdebatty14distance} was then used to measure the distance between two phoneme sequences.
The cons of using edit distance as a measure is the lack of offline indexing, so the similarity can only be calculated on the fly, preventing us from using a larger corpus.

\paragraph{Generative Approach.}
By treating transcription post-editing as a text-to-text translation problem, we created a generative machine-learning model that automatically translates a phoneme sequence to its reconstructed transcription.
We first converted all sentences into phoneme sequences using the g2pE package~\cite{g2pE2019} and used the tensorflow implementation~\cite{luong17} of LSTM~\cite{sutskever2014sequence} to learn the mapping between a phoneme sequence and its original sentences.
LSTM takes one phoneme sequence as input and can generate multiple outputs, which is suitable for our system.
Two large datasets, the Topical-Chat~\cite{Gopalakrishnan2019} and Switchboard Dialog Act Corpus~\cite{Jurafsky-etal:1997}, were used to train the model.
A total of 230,000 phoneme-to-transcription pairs were used as the training instances.
To prevent the model from generating repeated sentences such as ``There are three cars.'', ``There are 3 cars.'', and ``There are 3 cars!'', we normalized the data by removing punctuation and turning all numbers into Arabic numerals.
Additionally, two data augmentation strategies were used to diversify the generated transcription by {\em (i)} removing a phoneme with the probability of $0.1$, and {\em (ii)} exchanging a phoneme with a similar one with the probability of $0.1$.
The phoneme similarity is measured by the cosine similarity on phoneme embedding trained by word2vec~\cite{mikolov2013distributed}.
The augmentation is done five times and results in a total of 1.4 million instances for training. 
Table~\ref{tab:post-editing-result} shows that the sequence-to-sequence generative model produced alternative sentences that are better suited for the purpose of our system, which we used in the final version of \system.

\begin{table}[t]
    \centering
    \small
    \begin{tabular}{p{3.75cm}p{3.75cm}}
        \toprule \midrule
        \multicolumn{1}{c}{\textbf{Mistranscribed Sentence}} &  \multicolumn{1}{c}{\textbf{Post-editing Transcription}} \\ \hline
        \multicolumn{2}{c}{\textbf{Retrieval-based Model.}} \\ \hline
        What's your favorite \underline{seen} in the movie. & What's your favorite news? \\
        Not bad I have dinner \underline{was a} some friends. & And, I have a friend. \\
        Have you ever tried \underline{anyone a} sports before. & Have you ever served on a jury before? \\ \midrule
        
        \multicolumn{2}{c}{\textbf{Generation-based Model.}} \\ \hline
        \underline
        What's your favorite \underline{seen} in the movie. & What's your favorite in the movie? \\
        Not bad I have dinner \underline{was a} some friends. & Not bad i have always wanted some friends. \\
        Have you ever tried \underline{anyone a} sports before. & Have you ever tried any dunce caps before? \\ 
        \midrule \bottomrule
    \end{tabular}
    \caption{Result of the two post-editing speech transcription approaches. The retrieval-based model largely changes the input sentences; whereas the generation-based model modifies only a certain part of it. Therefore, the generation-based model is closer to our need of getting alternative transcriptions.}
    \label{tab:post-editing-result}
\end{table}

\subsection{Automatic Response Suggestion}
\label{sec:response-suggestion}
We used Cleverbot\footnote{Cleverbot: https://www.cleverbot.com/} to generate responses automatically.
Cleverbot is a third-party API that can generate generic responses for any arbitrary given sentences.
For each message and each of its post-edited versions, \system obtains at most five different automatic responses from Cleverbot.
Given one incoming message and three post-edited ones, \system obtains a total of $(3+1)\times5=20$ automatic responses.
The Cleverbot API has a response rate limitation of approximately one response per second, so generating all the responses will take around 20 seconds.
To prioritize the original transcription, \system first generates three responses for the original transcription, and then goes one by one for the alternative transcriptions.


%% file: 04_user_study.tex
To evaluate \system, we conducted a set of in-lab user study with 17 participants.
This study has been approved by the IRB office of the authors' institute.
Note that the study was conducted in late 2019, and we discussed the changes in Alexa since then in the Discussion Section.

\subsection{Pilot Study}
We conducted a pilot study to empirically explore the parameters and detailed setups of \system. 
In the study, one author talked to the Echo device (as the user) and a recruited participant operated an earlier version of \system in a separate room.
Both participants were instructed to hold casual social conversations.
Each session kept going for 7-15 turns, 
resulting in a transcript of around 20-30 sentences.
A total of seven sessions were tested, spanning 82 turns. 
We found that 
the speech transcripts provided by Alexa were not always correct, especially for phrases with similar pronunciation.
For example,
\say{Oh, can you describe the favorite city that you have ever been to?} being transcribed into \say{\textit{no} can you describe the favorite city \textit{let} you have ever been to?}.
We also observed that having the worker to click buttons is much faster than to type text. 
Of the 82 responses from the worker, 22 were chosen from Cleverbot options and 11 from default buttons.
For the 33 responses created by clicking buttons, the average delay was 11.8 seconds (SD = 4.2).
For the 49 worker-typed responses, the average delay was 17.7 seconds (SD = 4.97).

\subsection{Participants}
The participants for the user study were recruited through personal networks and flyers.
A total of 9 individuals were recruited as the users; 5 males and 4 female, aged between 18 and 35. 
For the workers, 8 individuals were recruited; 4 male, 3 female, and 1 prefer not to say; aged between 20 and 30.
All participants were undergraduate or graduate students in an university. Due to small scale recruiting, we can not eliminate the fact that workers and users know each other. Some participants are actually friends and intended to participate as a group. We came to notice that the conversation for these groups are mostly pretty fluid, and the topics tend to be more relatable, leading to a more enjoyable experience. U4 and W4 were roommates and U8, W8, and U9 were friends who signed up together, thus were paired up as a team to participate the study. U8 and U9 split up the normal user task in half and one would sit out while the other person was speaking, though observation and discussion were welcomed. 

\subsection{Study Setup and Protocol}

\begin{figure}[t]
    \centering
    \includegraphics[width=0.48\linewidth]{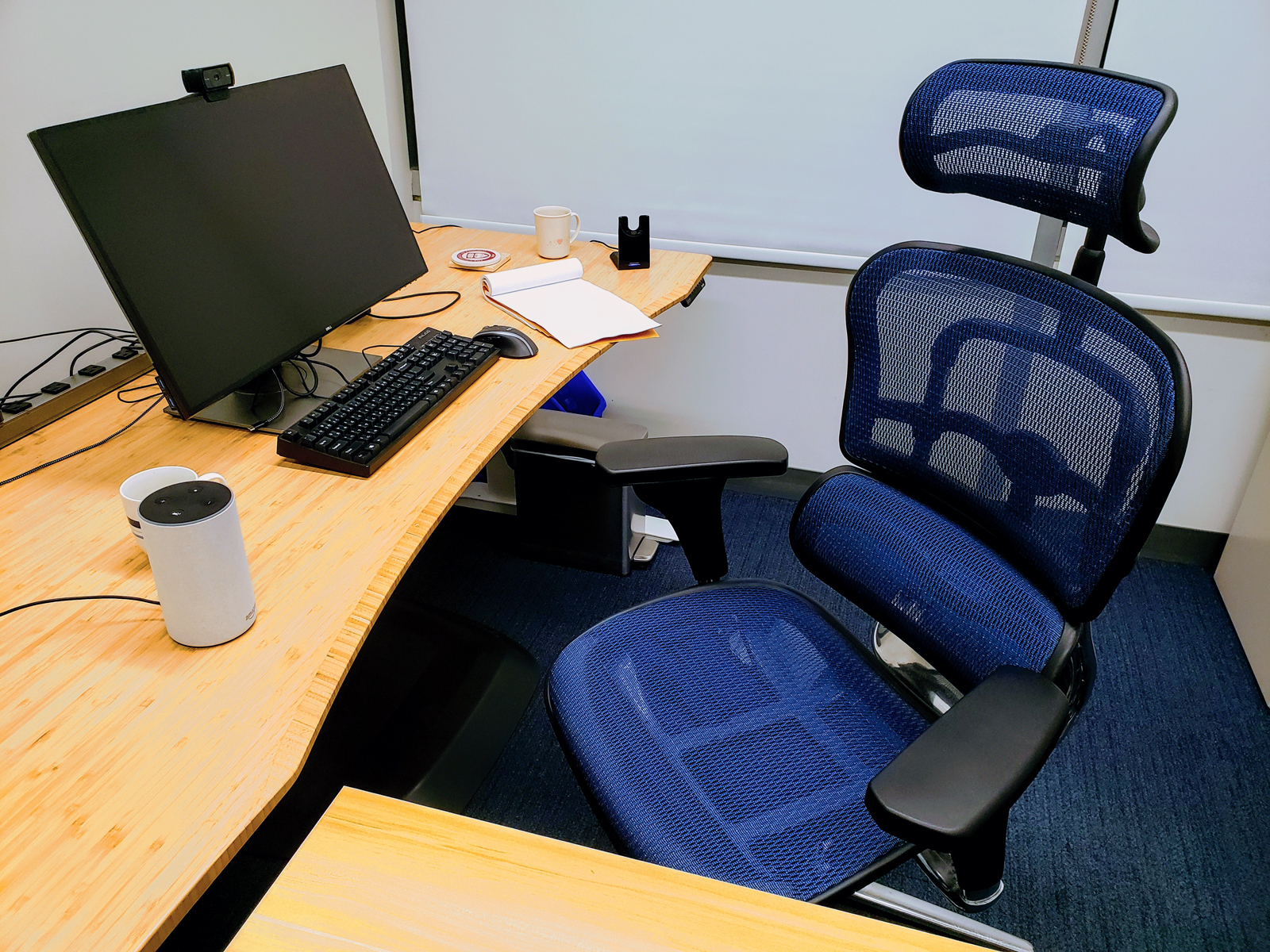}
    \hspace{0.1cm}
    \includegraphics[width=0.48\linewidth]{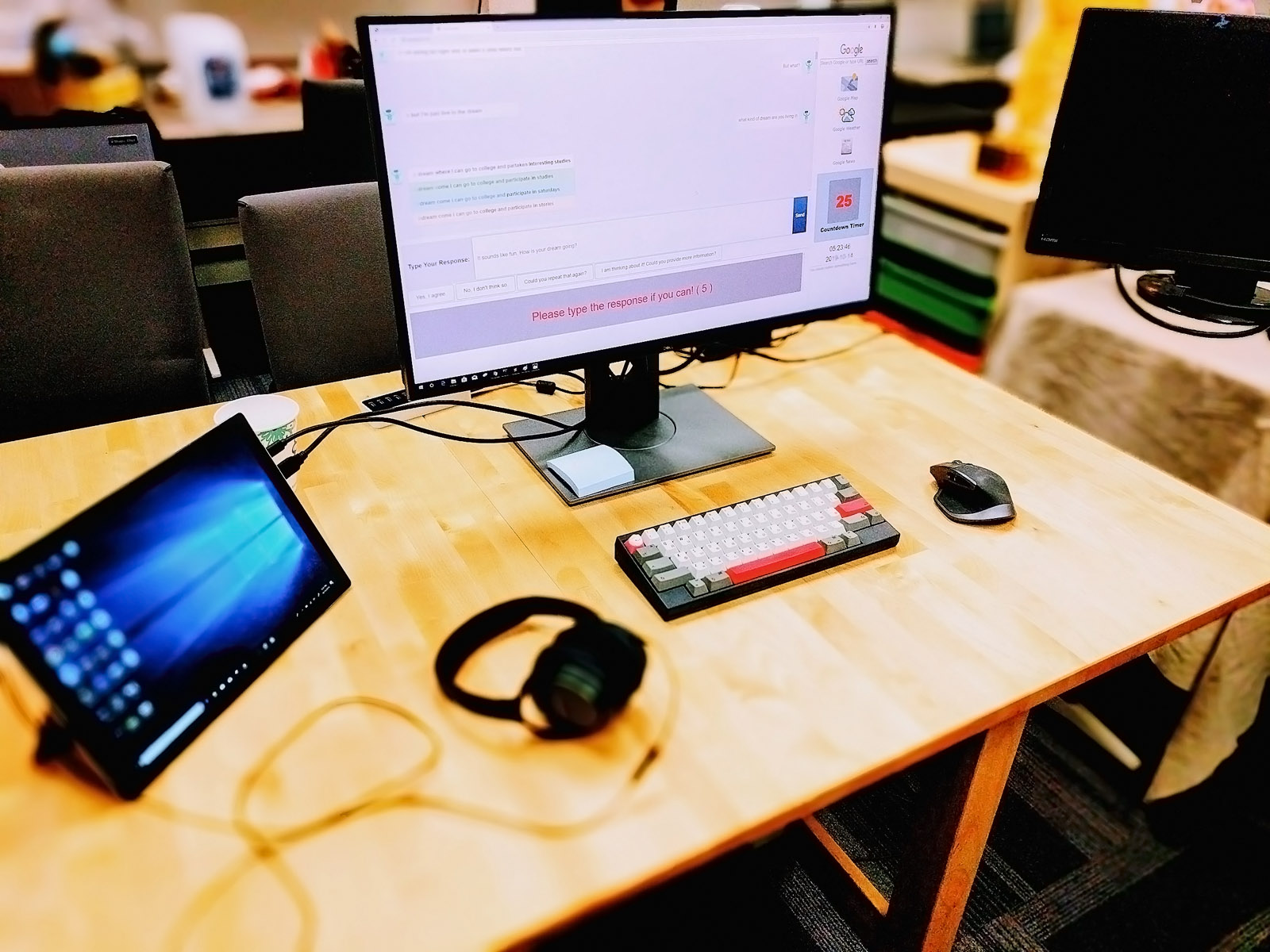}
    \caption{The user study station for the user (left) and the \worker (right). The user station is arranged in a private room with the Echo set up. The \worker station has a large monitor, a keyboard, a mouse, and a headphone set up to simulate a call center environment.}
    \vspace{-0.2cm}
    \label{fig:user-study-station}
\end{figure}

The user study involved two participants, one as the user and the other as the worker.
At the beginning of the study,
an author first introduced the concept of \system, explained the roles of both participants, and clarified the fact that they will be conversing with each other in the following study. The user and the worker were then assigned to two different rooms so direct communication was prohibited. 
User and \worker stations are shown in Figure~\ref{fig:user-study-station}.
At the user station, the user would first chat with two Alexa Prize Socialbots~\cite{alexaprize_2019} for five minutes each.  
It can help users get familiar with the interaction pattern of Echo devices and can be treated as baseline in the study.
Socialbots were randomly picked from one of the ten teams participating in the prize challenge by Amazon and was not revealed to outsiders. 
After interacting with the Alexa Prize Socialbots, users would be asked to chat with \system for 20 minutes.
Topics covered chit-chat and open-domain questions, but both sides were encouraged to chat freely without any kind of constraint on the topic.
In addition, in case of users run out of ideas, we provided around 100 questions, such as \say{What's the best way to learn a foreign language?}, 
in the form of a script.
At the \worker station, one author first taught the participant how to use the interface and had them get familiar with with the system for ten minutes.
And then, this participant served as a worker for 20 minutes using all the provided functionalities.
After the chat was over, users and \workers needed to fill up a questionnaire as an assessment for their experience with \system separately.
Finally, \$15 dollars of compensation will be given to both participants after the entire study was over.

\begin{figure}
  \centering
  \includegraphics[width=\linewidth]{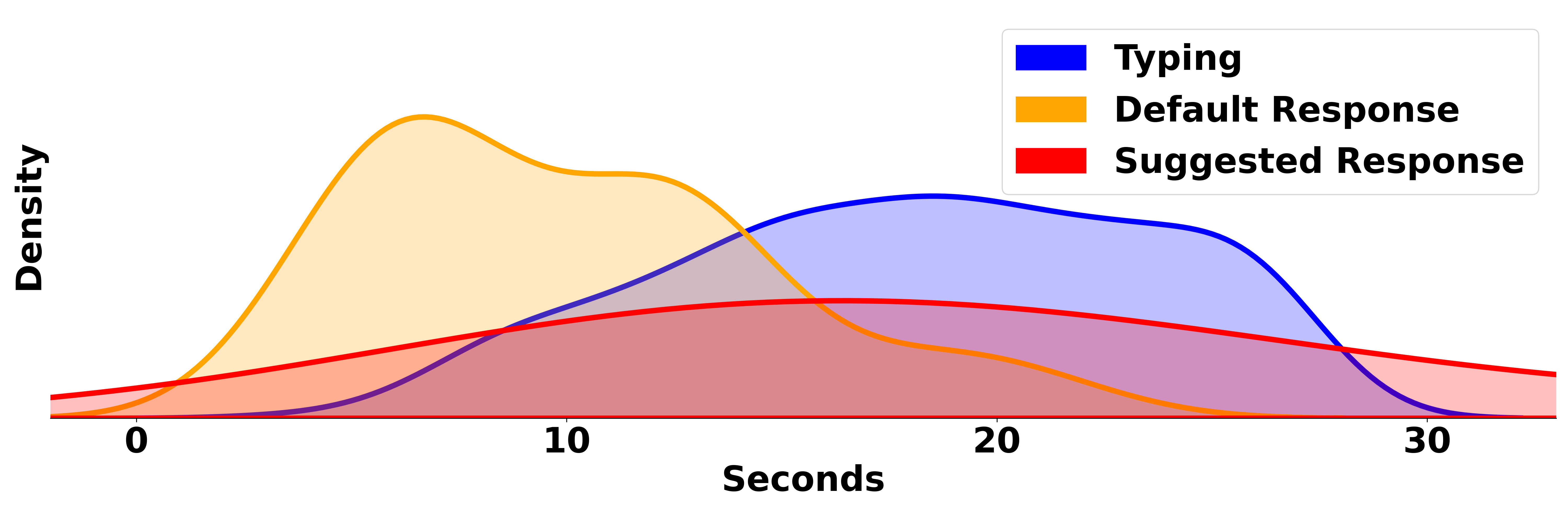}
  \caption{KDE distribution of the latency. (Note that KDE is an estimation of the distribution so exceeding 0-25 seconds is possible.) From the distribution, we found that {\em (i)} default response gives systematically lower latency; {\em (ii)} typing normally takes around 20 seconds.}
  \label{fig:distribution}
\end{figure}

\begin{figure}
  \centering
  \includegraphics[width=\linewidth]{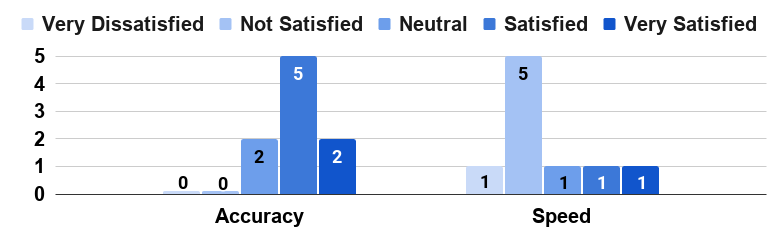}
  \caption{Users rating on ``Response \textbf{accuracy} to your questions'' and ``Response \textbf{speed} to your questions''. Most users agree that \system gives accurate response but is slow.}
  \label{fig:preference-alexa}
\end{figure}

\subsection{End-to-End System Latency}
Eight pairs of participants produced a total of 350 turns of conversation.
The latency is defined as the duration between the time when the system received the user message and when the system received the worker response.
The average latency is \textbf{17.68 seconds} (SD = 6.29). 
Among the 350 responses, 12 were responses generated by Cleverbot with an average latency of 22.5 seconds (SD = 13.16); and 
27 were the default responses that sent by simply clicking the pre-defined button such as ``Yes, I agree.'', with an average latency of 10.04 seconds (SD = 4.48).
Figure~\ref{fig:distribution} shows that the default response gives systematically lower latency, and typing on average takes around 20 seconds. 




\subsection{Conversation Quality}


After each experiment trial, we had the participants to fill a questionnaire survey to assess their user experience.
We asked the users (N = 9, not workers) to directly rate the overall quality of the system on a 5-point Likert scale, from ``Low Quality'' (1) to ``High Quality'' (5).
The average score is 3.67.
We also asked users: ``Compared with the automated Alexa Socialbots that you just tried, how would you describe our system?'', with a 5-point Liker scale, from ``Much Worse'' (1) to ``Much Better'' (5).
From the results, 3 participants thought our system is ``Better'', and 2 thought our system is ``Much Better''.
And 3 participants considered our system are similarly good as automated Alexa Socialbots.
Only 1 participant rated our system as ``Worse''.


\subsection{User's Satisfaction Level}
We further asked users to rate their satisfaction levels on the {\em (i)} accuracy and {\em (ii)} speed using in a 5-point Likert scale, from ``Very Dissatisfied'' (1) to ``Very satisfied'' (5).
The average score of response quality was 4.0, but that of response speed was only 2.56 (Figure~\ref{fig:preference-alexa}).
Namely, \textbf{users were satisfied with the quality of the conversation, but strongly dissatisfied with the turnaround time of \system.}
Participants expressed a similar sentiment in their feedback.
For example, U1 said: ``I was impressed with the detail of many of the answers.''
U8 also said: ``It's slower and less consistent, but it can give better answers depending on the question." 
U9 compared \system with Alexa's Socialbots and said: "(\system is) More consistent but was definitely slower.''
We concluded that even with the current setup and features of \system, which can limit workers' ability to provide optimum responses human workers still outperform automated agents.

\subsection{Conversation Examples} 
As an example, we show the complete log between the participant U8 (user end) and W8 (worker end).
During the conversation,
the worker used the search pane to quickly get the information to fulfill user's need ({\em e.g.,} ``how many bones are in my hand'') and asked questions for clarification when built-in ASR failed ({\em e.g.,} ``Did you mean EAST COAST?''). Identifiable information in the example such as location and names are replaced with placeholders.

\paragraph{Conversation 1 (U8-W8)}
\begin{choruschat}
    \user{how many bones are in my hand }
    \echo{THere are 27 bones in your hand}
    \user{how do I sign recursion to a 5 year-old}
    \echo{Peat and Repeat were in a boat, peat fell out, who was left?}
    \user{what movies are playing tonight}
    \echo{Probably Joker}
    \user{what's what is the score of the packers game}
    \echo{It hasn't changed since I told you last}
    \user{what}
    \echo{Could you repeat that again?}
    \user{how many planets are in the solar system}
    \echo{9 Cause I believe Pluto counts}
    \user{is a hot dog a sandwich}
    \echo{No, but one can argue that all matter is a sandwich}
    \user{can you explain more on that topic}
    \echo{If a sandwich is nothing but an item surrounded by other items that is an atoim}
    \user{trump be impeached}
    \echo{Yes, Trump is being impeached}
    \user{find me a restaurant near-by preferred lie Italian food}
    \echo{I think there's a \{RESTAURANT\} downtown}
    \user{what's the best type of cheese}
    \echo{Parmigianno Reggiano is known as the king of cheeses}
    \user{coffee or tea}
    \echo{I prefer cold water for my CPU}
    \user{what is the west coast to best coast}
    \echo{Did you mean "EAST COAST"?}
    \user{how many bones are in my foot}
    \echo{There are 26 bones in a foot but I can make there be more}
    \user{how far is it to \{SCHOOL\}}
    \echo{\{SCHOOL\} is about 3 hours away, or 5 if you stop at the \{LOCATION\}}
    \user{what's going out of the San Diego zoo right now}
    \echo{I am thinking about it! Could you provide more information?}
    \user{sure what is going on in the month of October at the San Diego zoo located in california}
    \echo{You could look at a lion}
\end{choruschat}

Another example shows the partial log from the participants U4 (user end) and W4 (worker end).
Note that U4 and W4 were roommates as stated in our recruiting process.
It is not easy to have such a fluent and coherent conversation using \system.
Thanks to the fact that both the user and worker share the same background and context, they enjoyed \system a lot.

\paragraph{Conversation 2 (U4-W4)}
\begin{choruschat}
    \user{hey how's billy doing}
    \echo{He looks very bored but doing great!}
    \user{no he's so cute tell him he's a good boy}
    \echo{I told him and fed him. Now, he needs a walk.}
    \user{you did you not bring any toys you need toys to play with}
    \echo{I should have brought his toy so he could play with it. sorry bailey}
    \user{yeah you should be sorry now you're making I'm bored}
    \echo{He will be fine.}
    \user{do you consider yourself a dog person}
    \echo{yes but i also love cat. can i be dog slash cat person?}
    \user{no you have to pick 1}
    \echo{you are giving me a hard time}
    \user{me there you can pick neither 2}
    \echo{I will pick dog person then}
    \user{why}
    \echo{they are friendly and loyal unlike cats they are mean}
    \user{so then you're not really a cat and dog person you just throw cats in there just a seem like you're a good person}
    \echo{I think cats are cute when they are mean because i see myself in that}
    \user{you know that doesn't even surprise me and it makes me really mad have to say that}
    \echo{why are you mad? dont be mad! Anger can hurt you.}
    \user{it can also help me grow and become stronger}
    \echo{That is true. Well, I hope I help you grow and become st}
    \user{who is ST}
    \echo{i mean to say That is true. Well, I hope I help you grow and become stronger}
    \user{oh okay sorry I was confused her for a bit}
    \echo{I was not able to finish typing in time}
    \user{do you prefer friends were like cats or dogs}
    \echo{dogs of course because i can be the only cat}
    \user{so does that make me a dog Ben}
    \echo{Yes you are my toy}
    \user{what}
    \echo{\{NAME\} is my favorite toy}
    \user{what type of dog am I}
    \echo{I think you are like a German Shepherd}
    \user{why is your name shepherd}
    \echo{what do you mean? My name is not shepherd}
    \user{I didn't say that I said why do you think I'm a German shepherd}
    \echo{Because you are like a guard dog and bark when you need}
    \user{do you know that's that's very far okay I knew you're gonna say something like that too}
    \echo{You are like the  shepherd at the airport to check drugs!}
    \user{that's really funny}
    \echo{If you smell something you bark something}
    \user{okay do you wanna watch Netflix after this}
    \echo{I would love to but i am going to take graduation pictures after this}
    \user{oh you're such a loser}
    \echo{I am hungry}
    \user{oh okay that's nice to know}
    \echo{I can't eat anything. I am meeting up with my friends at 1pm}
    \user{I do that actually sucks why do you keep taking grad photos you've taken like over a thousand and now}
    \echo{Because i got friends unlike you hahaha}
    \user{you're so mean}
    \echo{remember  I am the only cat}
    \user{that's true I'll just watched enchanted without you then}
    \echo{go ahead. have fun. dont be a spoiler}
    \user{all spoil everything}
    \echo{i will buy a 3M ear bug}
    
    \voiceover{SKILL BREAKS DUE TO USER ACCIDENTALLY TERMINATE \system. USER REACTIVATE \system IMMEDIATELY}
    
    
    \user{sorry I had a go away for a second but I'm back}
    \echo{Hi welcome back}
    \user{was that a dad joke}
    \echo{I think your dad would love my joke}
    \user{I don't think he would my dad's been kind of mad at you lately for coughing and stuff}
    \echo{Are you seriously?}
    \user{yes I'm seriously I'm seriously Sirius}
    \echo{Ok tell him I am sorry but you got sick first you gave the sickness to me}
    \user{hey how much wood could a woodchuck chuck if a woodchuck could chuck wood}
    \echo{I think 300 pound sof woods}
    \user{like pine trees}
    \echo{I think 300 pounds of woods}
    \user{no how's your come up with that number}
    \echo{i made it up}
    \user{okay that you know solid facts}
    \echo{New York state wildlife expert Richard Thomas found that a woodchuck could (and does) chuck around 35 cubic feet of dirt in the course of digging a burrow. Thomas reasoned that if a woodchuck could chuck wood, he would chuck an amount equivalent to the weight of the dirt, or 700 pounds.}
    \user{is that true or did you just make that up again}
    \echo{According to google}
    \user{he can I be a woodchuck than instead of a German shepherd}
    \echo{No. Can we change the subject?}
    \user{sure what do you wanna talk about}
    \echo{Talk about where do you want to go if you travel to \{LOCATION\}}
    \user{I said I wanna go everywhere I wanna do everything}
    \echo{do you want to try durians?}
    \user{usher I'll try it although I heard they're really smelly}
    \echo{they are so good. One of my favorite. \ }
    \user{is your mom so gonna make that duck dish}
    \echo{I don't remember which duck dish you are talking about}
\end{choruschat}

%% file: 05_discussion.tex
Our user study demonstrated the superior and trade-offs of \system.
In this section, we discuss topics that are broader than the scope of our experiments.

\subsection{Workers' Challenges: Extremely Short Response Time}
\label{subsec:short-response-time}
We found that the critical factor for the \workers is the time constraint, which has clear trade-offs with the response quality.
In \system, we set a 25-second time constraint for the \workers to adapt for the short response time of voice-enabled devices.
This 25-second time constraint is based on the average of our pilot study's latency and average response time of an instant message~\cite{avrahami2008waiting}.
However, it was difficult for \workers to answer questions or to respond properly within such a short period of time. 
As one participant (W2) put in the survey: \say{It took me longer to search for specific information that required google search. For example: a certain name of a restaurant, a specific title of a movie.}
W7 also said: \say{..., if the timer could be increased, it would have been much helpful!}
W1 had a similar comment: \say{..., the response time could be a bit short. Thus, I could only reply short answers.}

We observed in a few cases where the \worker choose to type a short response, such as \say{I don't know}, instead of providing detailed answers because they did not have enough time.
Furthermore, if the question is out of \workers' knowledge domain, it could take much longer for \workers to answer it. 
For example, one of the worker's (W1) feedback is: \say{Also, sometimes users could say something that workers don't know how to response, such as ``give me a history quiz'' or ``you were giving me there any short answers could you please give me longer answers''. The question itself could be very general or out of workers' knowledge domain. It may be hard to give an appropriate answer within the time constraint. Therefore, the time constraint affects the quality of the response.}
In contrast, when the \worker was familiar with questions, the reply speed and accuracy were much better.
For example, W8 said: \say{...but in circumstances where I already knew the information I was able to give a more informed response than a simple google search.}

Note that our user studies were conducted between September 10, 2019 to September 14, 2019, when Alexa's timeout were longer.
We are aware of that the current timeout of Alexa (as of July 2, 2021) is shortened to roughly 8 seconds.
To the best of our knowledge, this timeout is not configurable or negotiable, making it even more difficult for \workers to respond effectively.

\subsection{Workers' Needs for Better Search Supports}
We asked \workers to rate the usefulness of each features in a 5-point Likert scale, from Very Useless (1) to Very Useful (5).
Results show that, \workers on average consider the web search quick links (4.125) and static response button (3.625) are useful, and automatic suggested responses mildly useful (3.125).
We also asked \workers to rate their satisfaction level toward the accuracy of the two features we created, 
the alternative transcripts generation and automatic response suggestion,
on a 5-point Likert scale, from Very Dissatisfied (1) to Very satisfied (5).
Automatic response suggestion received an average of 3.375 accuracy rating, while alternative transcripts generation was only rated 2.625.
In general, the workers gave us positive feedback of these automated support features. 

We also asked \workers: ``If you could improve this system, what function(s) would you add?''
Most suggestions were about better supporting web searches.
For example, the \worker W4 mentioned: ``If it is for asking facts, the system can automatically search on google and have several suggested answers popped out for users to choose.''
The \worker W6 also have a similar suggestion: ``When being asked about questions regarding directions, food places, weather, etc... the system starts searching automatically rather than me manually typing in.''
In addition, current \system opens a new browser tab for Google Search, and many \workers suggested that the search page should be embedded in the worker interface.

\subsection{Users' Challenges: Cut-Offs of Conversations}

A common problem which occurred in the user study is cut-offs of conversations.
When the user paused too long between two words during a speech, the device sometimes stopped listening and sent out the utterances even when the sentences were not yet finished.
Almost all users mentioned this problem in the survey and hoped it could be improved.
For example, 
\say{I would add increased response time, and the ability to interrupt the the device mid-response. This will make it feel more human-like} (U1)
or, 
\say{Alexa sometimes cut me off mid sentence when I had a gap in speech. A system to improve that would be helpful.} (U3).

We also noticed that familiarity with the current interaction pattern of Echo devices and whether the user felt comfortable talking to a device significantly affect the user's ability to deal with cut-offs.
Due to the turn taking design of Echo, users' speech will only be taken in while the light ring is on.
After the light dimmed, any further speech will not be transcribed, resulting in a cut-off constraint on the users' end.
Cut-offs mostly happen in the middle of a sentence, often caused by a longer pause between two words, which the device will consider the speech to be over.
Another kind of cut-off happen at the start of a speech.
After Alexa finish reading out the message from the worker, it will enter the listening mode and start to transcribe.
If the user does not start to talk within a period of time, it will either try to transcribe any background noise picked up by the device or turn off the skill completely.
For participants who were more familiar with or accustomed to interact with Alexa, the conversation will proceed more smoothly, meaning they encountered less cut-offs comparing to other users.
Some participants expected Echo to wait until they start to speak or be listening to their speech all the time.

\subsection{Reasons for Having Humans in the Loop}
Despite being unsatisfied about the long latency,
according to the user survey, 
most participants still considered \system performed better than automated Alexa Socialbots.
From the \workers' feedback, we can further gain some interesting insights.
For example, one \worker said, \say{...there was one question: okay if X plus 7 equals 15. The question itself was incomplete, it could be caused by flaws in voice detection or user end. However, workers would intuitively understand the question ``X plus 7 equals 15. What is X?'', and provide the answer ``X = 8''.}
Compared with automated approaches, human workers are better at guessing or even predicting the user intent by making sense out of contextual information and prior conversations and can produce better responses.

We are aware of several common arguments for having human-powered components inside automated systems.
The \textbf{social-technical gap argument} suggests that humans' extraordinary task capability can be used in automated systems to reduce the social-technical gap~\cite{ackerman2000intellectual} between ``what people want the computers to do'' (converse naturally) and ``what the computers can practically do (limited voice command.)''  
This Wizard-of-Oz-like approach allows researchers to study the trade-offs and user scenarios of near-future intelligent systems that can not be fully automated yet~\cite{huang2019automating}.
The \textbf{long tail argument} suggests that the real-world data often follows a long-tailed distribution ({\em e.g.}, topics that people want to talk about), where computers can solve common, frequent cases and assign the rarer, more unique cases to humans~\cite{chang2017revolt}.
Finally, the \textbf{reality check argument} emphasizes the imperfection of existing technologies, which necessitates human oversight in many real-world scenarios.

Our study results echoed some of these arguments. 
However, challenges can arise from the fact that many existing infrastructures are not created to work with human-powered components, making it expensive or extra hard to honor the promised benefits.

%% file: next-step.tex

Our user study explicates the steep challenges of incorporating humans into the loop of automated systems that require extremely short response times.
Here, we submit three possible future directions that might mitigate the hardship, 
with the goal of helping future explorations.

\begin{itemize}
    \item \textbf{Creating smart speakers that allows voice-based back ends.}
\system adopts a voice-to-text approach where the unbalanced time demands on the voice and text side makes it a problem.
Compared to such scenario, a voice-to-voice approach is closer to the general conversational pattern.
For example, Pechat\footnote{Pechat: https://pechat.jp}, a gadget made for children under six, can be placed on stuffed animals pretending that the stuffed animals are capable of talking. 
Parents can talk and interact with their child through Pechat via the mobile app.
The voice-to-voice scenario largely reduces the negative effects caused by the voice-to-text setting.
In other words, this direction falls back to traditional voice-based call center-like set ups.

    \item \textbf{Moving toward highly-specified, task-oriented conversations.}
Parts of the human latency are caused by the large size of search space of open conversations.
If \system moves away from the open conversation and focuses on a well-defined domain, the latency could be significantly shortened.
In this case, well-trained receptionists accompanied with sufficient knowledge base, familiarity, and tools specifically for the task will help users more seamlessly. 
On the other hand, the more grounded conversation also makes the content more predictable and, therefore, more likely for the system to automatically generate responses in buttons, which also helps shorten the delay.    


    \item \textbf{Making asynchronous dialogues possible.}
To tackle the core challenge set forth by extremely short response time, one possibility is to allow
asynchronous conversations between \system and the user, which
will free the system from the traditional synchronous interaction modality.
The user initiates a discussion with \system by simply talking to it, but without expecting an immediate response.
Behind the scenes, \system employs both AI and human computation workflows to explore the question, collect relevant information, brainstorm and reason follow-up topics, and decide how to respond.
When ready, the system reaches out to the user via multiple channels, including voice, emails, instant messaging applications, or even social media, to continue the conversation.
The user can respond to the system via their preferred channel whenever they want, and \system will use the response to develop the conversation further.


\end{itemize}





%% file: 06_conclusion.tex
This paper introduces \system, a prototype system that allows a human worker to converse with the user synchronously via an Amazon's Echo device without access to the user's voice recordings.
Through the user studies, we showed that it is feasible and beneficial-- but challenging-- to combine smart speakers and human intelligence.
We observed that the main challenge for users is the cut-offs of ongoing conversations; 
and the main challenge for \workers is the extremely short response time.
We also found that the \workers considered the web search support is the most useful feature.
In our user study, many users expressed their frustration of the long latency of the system.
However, it is noteworthy that we conducted our user study only under a single condition: users were sitting in a quite conference room. 
We are aware of the fact that the acceptable response latency varies among different user scenarios.
For example, in moving vehicles, since drivers' attention will mainly be on the road, either drivers need the information as soon as possible, such as asking for directions while lost, or they can wait for the response, such as chatting or music search. 
It requires more research to fully understand the proper latency for voice-enable devices in different scenes.


Our work explores the possibilities and challenges of human‐in‐the‐loop smart speakers, informing the designs of future systems facing various real‐world constraints.

\section{Acknowledgements}
We thank Ming-Ju Li, Tiffany Knearem, Sooyeon Lee, and Namo Asavisanu for their valuable help and feedback. We also thank the participants who participated in our studies.

%% file: main.bbl
\begin{thebibliography}{41}
\providecommand{\natexlab}[1]{#1}
\providecommand{\url}[1]{\texttt{#1}}
\providecommand{\urlprefix}{URL }
\expandafter\ifx\csname urlstyle\endcsname\relax
  \providecommand{\doi}[1]{doi:\discretionary{}{}{}#1}\else
  \providecommand{\doi}{doi:\discretionary{}{}{}\begingroup
  \urlstyle{rm}\Url}\fi

\bibitem[{ale(2019)}]{alexaprize_2019}
 2019.
\newblock alexaprize.
\newblock \urlprefix\url{https://developer.amazon.com/alexaprize}.

\bibitem[{Ackerman(2000)}]{ackerman2000intellectual}
Ackerman, M.~S. 2000.
\newblock The intellectual challenge of CSCW: the gap between social
  requirements and technical feasibility.
\newblock \emph{Human--Computer Interaction} 15(2-3): 179--203.

\bibitem[{Adam(2018)}]{catchall}
Adam, S. 2018.
\newblock Amazon Alexa: store user's words.
\newblock
  \urlprefix\url{https://stackoverflow.com/questions/37249475/amazon-alexa-store-users-words/53334157#53334157}.

\bibitem[{Avrahami, Fussell, and Hudson(2008)}]{avrahami2008waiting}
Avrahami, D.; Fussell, S.~R.; and Hudson, S.~E. 2008.
\newblock IM waiting: timing and responsiveness in semi-synchronous
  communication.
\newblock In \emph{Proceedings of the 2008 ACM conference on Computer supported
  cooperative work}, 285--294. ACM.

\bibitem[{Bernstein et~al.(2011)Bernstein, Brandt, Miller, and
  Karger}]{bernstein2011crowds}
Bernstein, M.~S.; Brandt, J.; Miller, R.~C.; and Karger, D.~R. 2011.
\newblock Crowds in two seconds: Enabling realtime crowd-powered interfaces.
\newblock In \emph{Proceedings of the 24th annual ACM symposium on User
  interface software and technology}, 33--42. ACM.

\bibitem[{Bigham et~al.(2010)Bigham, Jayant, Ji, Little, Miller, Miller,
  Miller, Tatarowicz, White, White et~al.}]{bigham2010vizwiz}
Bigham, J.~P.; Jayant, C.; Ji, H.; Little, G.; Miller, A.; Miller, R.~C.;
  Miller, R.; Tatarowicz, A.; White, B.; White, S.; et~al. 2010.
\newblock VizWiz: nearly real-time answers to visual questions.
\newblock In \emph{Proceedings of the 23nd annual ACM symposium on User
  interface software and technology}, 333--342. ACM.

\bibitem[{Blair and Abdullah(2019)}]{10.1145/3311957.3359487}
Blair, J.; and Abdullah, S. 2019.
\newblock Understanding the Needs and Challenges of Using Conversational Agents
  for Deaf Older Adults.
\newblock In \emph{Conference Companion Publication of the 2019 on Computer
  Supported Cooperative Work and Social Computing}, CSCW ’19, 161–165. New
  York, NY, USA: Association for Computing Machinery.
\newblock ISBN 9781450366922.
\newblock \doi{10.1145/3311957.3359487}.
\newblock \urlprefix\url{https://doi.org/10.1145/3311957.3359487}.

\bibitem[{Bohn(2019)}]{bohn_2019}
Bohn, D. 2019.
\newblock Exclusive: Amazon says 100 million Alexa devices have been sold.
\newblock
  \urlprefix\url{https://www.theverge.com/2019/1/4/18168565/amazon-alexa-devices-how-many-sold-number-100-\\million-dave-limp}.

\bibitem[{Braun et~al.(2019)Braun, Mainz, Chadowitz, Pfleging, and
  Alt}]{Braun:2019:YSD:3290605.3300270}
Braun, M.; Mainz, A.; Chadowitz, R.; Pfleging, B.; and Alt, F. 2019.
\newblock At Your Service: Designing Voice Assistant Personalities to Improve
  Automotive User Interfaces.
\newblock In \emph{Proceedings of the 2019 CHI Conference on Human Factors in
  Computing Systems}, CHI '19, 40:1--40:11. New York, NY, USA: ACM.
\newblock ISBN 978-1-4503-5970-2.
\newblock \doi{10.1145/3290605.3300270}.
\newblock \urlprefix\url{http://doi.acm.org/10.1145/3290605.3300270}.

\bibitem[{Chang, Amershi, and Kamar(2017)}]{chang2017revolt}
Chang, J.~C.; Amershi, S.; and Kamar, E. 2017.
\newblock Revolt: Collaborative crowdsourcing for labeling machine learning
  datasets.
\newblock In \emph{Proceedings of the 2017 CHI Conference on Human Factors in
  Computing Systems}, 2334--2346. ACM.

\bibitem[{Cheng et~al.(2018)Cheng, Yen, Chen, Chen, and
  Hiniker}]{10.1145/3202185.3202749}
Cheng, Y.; Yen, K.; Chen, Y.; Chen, S.; and Hiniker, A. 2018.
\newblock Why Doesn’t It Work? Voice-Driven Interfaces and Young Children’s
  Communication Repair Strategies.
\newblock In \emph{Proceedings of the 17th ACM Conference on Interaction Design
  and Children}, IDC ’18, 337–348. New York, NY, USA: Association for
  Computing Machinery.
\newblock ISBN 9781450351522.
\newblock \doi{10.1145/3202185.3202749}.
\newblock \urlprefix\url{https://doi.org/10.1145/3202185.3202749}.

\bibitem[{Choi et~al.(2018)Choi, Kim, Han, Son, and
  Cho}]{Choi:2018:DCV:3267305.3267638}
Choi, S.; Kim, Y.; Han, E.; Son, J.; and Cho, J. 2018.
\newblock Designing Conversational Voice User Interface for Improving Intimacy
  of Shared Invehicle.
\newblock In \emph{Proceedings of the 2018 ACM International Joint Conference
  and 2018 International Symposium on Pervasive and Ubiquitous Computing and
  Wearable Computers}, UbiComp '18, 33--37. New York, NY, USA: ACM.
\newblock ISBN 978-1-4503-5966-5.
\newblock \doi{10.1145/3267305.3267638}.
\newblock \urlprefix\url{http://doi.acm.org/10.1145/3267305.3267638}.

\bibitem[{Debatty(2014)}]{tdebatty14distance}
Debatty, T. 2014.
\newblock java-string-similarity.
\newblock
  \urlprefix\url{https://github.com/tdebatty/java-string-similarity\#normalized-metric-similarity-and-distance}.

\bibitem[{Gopalakrishnan et~al.(2019)Gopalakrishnan, Hedayatnia, Chen,
  Gottardi, Kwatra, Venkatesh, Gabriel, and Hakkani-Tür}]{Gopalakrishnan2019}
Gopalakrishnan, K.; Hedayatnia, B.; Chen, Q.; Gottardi, A.; Kwatra, S.;
  Venkatesh, A.; Gabriel, R.; and Hakkani-Tür, D. 2019.
\newblock {Topical-Chat: Towards Knowledge-Grounded Open-Domain Conversations}.
\newblock In \emph{Proc. Interspeech 2019}, 1891--1895.
\newblock \doi{10.21437/Interspeech.2019-3079}.
\newblock \urlprefix\url{http://dx.doi.org/10.21437/Interspeech.2019-3079}.

\bibitem[{Guo et~al.(2018)Guo, Jain, Ghose, Laput, Harrison, and
  Bigham}]{10.1145/3264921}
Guo, A.; Jain, A.; Ghose, S.; Laput, G.; Harrison, C.; and Bigham, J.~P. 2018.
\newblock Crowd-AI Camera Sensing in the Real World.
\newblock \emph{Proc. ACM Interact. Mob. Wearable Ubiquitous Technol.} 2(3).
\newblock \doi{10.1145/3264921}.
\newblock \urlprefix\url{https://doi.org/10.1145/3264921}.

\bibitem[{Huang et~al.(2022)Huang, Huang, Deng, Shen, Kuan, and
  Huang}]{huang2022too}
Huang, S.-H.; Huang, C.-Y.; Deng, Y.; Shen, H.; Kuan, S.-C.; and Huang,
  T.-H.~K. 2022.
\newblock Too Slow to Be Useful? On Incorporating Humans in the Loop of Smart
  Speakers.
\newblock In \emph{Works-in-Progress Papers and Demonstration Papers of the
  AAAI Conference on Human Computation (HCOMP 2022)., HCOMP}.

\bibitem[{Huang(2019)}]{huang2019automating}
Huang, T.-H. 2019.
\newblock On Automating Conversations.
\newblock \emph{arXiv preprint arXiv:1910.09621} .

\bibitem[{Huang, Chen, and Bigham(2017)}]{huang2017real}
Huang, T.-H.; Chen, Y.-N.; and Bigham, J.~P. 2017.
\newblock Real-time on-demand crowd-powered entity extraction.
\newblock \emph{arXiv preprint arXiv:1704.03627} .

\bibitem[{Huang, Chang, and Bigham(2018)}]{huang2018evorus}
Huang, T.-H.~K.; Chang, J.~C.; and Bigham, J.~P. 2018.
\newblock Evorus: A Crowd-powered Conversational Assistant Built to Automate
  Itself Over Time.
\newblock In \emph{Proceedings of the 2018 CHI Conference on Human Factors in
  Computing Systems}, 295. ACM.

\bibitem[{Huang et~al.(2016)Huang, Lasecki, Azaria, and
  Bigham}]{huang2016there}
Huang, T.-H.~K.; Lasecki, W.~S.; Azaria, A.; and Bigham, J.~P. 2016.
\newblock " Is There Anything Else I Can Help You With?" Challenges in
  Deploying an On-Demand Crowd-Powered Conversational Agent.
\newblock In \emph{Fourth AAAI Conference on Human Computation and
  Crowdsourcing}.

\bibitem[{Jurafsky, Shriberg, and Biasca(1997)}]{Jurafsky-etal:1997}
Jurafsky, D.; Shriberg, E.; and Biasca, D. 1997.
\newblock Switchboard {SWBD}-{DAMSL} Shallow-Discourse-Function Annotation
  Coders Manual, Draft 13.
\newblock Technical Report 97-02, University of Colorado, Boulder Institute of
  Cognitive Science, Boulder, CO.

\bibitem[{Koni et~al.(2021)Koni, Al-Absi, Saparmammedovich, and
  Lee}]{10.1007/978-3-030-68452-5_39}
Koni, Y.~J.; Al-Absi, M.~A.; Saparmammedovich, S.~A.; and Lee, H.~J. 2021.
\newblock AI-Based Voice Assistants Technology Comparison in Term of
  Conversational and Response Time.
\newblock In Singh, M.; Kang, D.-K.; Lee, J.-H.; Tiwary, U.~S.; Singh, D.; and
  Chung, W.-Y., eds., \emph{Intelligent Human Computer Interaction}, 370--379.
  Cham: Springer International Publishing.
\newblock ISBN 978-3-030-68452-5.

\bibitem[{Laput et~al.(2015)Laput, Lasecki, Wiese, Xiao, Bigham, and
  Harrison}]{laput2015zensors}
Laput, G.; Lasecki, W.~S.; Wiese, J.; Xiao, R.; Bigham, J.~P.; and Harrison, C.
  2015.
\newblock Zensors: Adaptive, rapidly deployable, human-intelligent sensor
  feeds.
\newblock In \emph{Proceedings of the 33rd Annual ACM Conference on Human
  Factors in Computing Systems}, 1935--1944. ACM.

\bibitem[{Lasecki et~al.(2012)Lasecki, Miller, Sadilek, Abumoussa, Borrello,
  Kushalnagar, and Bigham}]{lasecki2012real}
Lasecki, W.; Miller, C.; Sadilek, A.; Abumoussa, A.; Borrello, D.; Kushalnagar,
  R.; and Bigham, J. 2012.
\newblock Real-time captioning by groups of non-experts.
\newblock In \emph{Proceedings of the 25th annual ACM symposium on User
  interface software and technology}, 23--34. ACM.

\bibitem[{Lasecki et~al.(2013)Lasecki, Wesley, Nichols, Kulkarni, Allen, and
  Bigham}]{lasecki2013chorus}
Lasecki, W.~S.; Wesley, R.; Nichols, J.; Kulkarni, A.; Allen, J.~F.; and
  Bigham, J.~P. 2013.
\newblock Chorus: a crowd-powered conversational assistant.
\newblock In \emph{Proceedings of the 26th annual ACM symposium on User
  interface software and technology}, 151--162. ACM.

\bibitem[{Levenshtein(1966)}]{levenshtein1966binary}
Levenshtein, V.~I. 1966.
\newblock Binary codes capable of correcting deletions, insertions, and
  reversals.
\newblock In \emph{Soviet physics doklady}, volume~10, 707--710.

\bibitem[{Lovato and Piper(2015)}]{Lovato:2015:STY:2771839.2771910}
Lovato, S.; and Piper, A.~M. 2015.
\newblock "Siri, is This You?": Understanding Young Children's Interactions
  with Voice Input Systems.
\newblock In \emph{Proceedings of the 14th International Conference on
  Interaction Design and Children}, IDC '15, 335--338. New York, NY, USA: ACM.
\newblock ISBN 978-1-4503-3590-4.
\newblock \doi{10.1145/2771839.2771910}.
\newblock \urlprefix\url{http://doi.acm.org/10.1145/2771839.2771910}.

\bibitem[{Lundgard et~al.(2018)Lundgard, Yang, Foster, and
  Lasecki}]{lundgard2018bolt}
Lundgard, A.; Yang, Y.; Foster, M.~L.; and Lasecki, W.~S. 2018.
\newblock Bolt: Instantaneous crowdsourcing via just-in-time training.
\newblock In \emph{Proceedings of the 2018 CHI Conference on Human Factors in
  Computing Systems}, 1--7.

\bibitem[{Luong, Brevdo, and Zhao(2017)}]{luong17}
Luong, M.; Brevdo, E.; and Zhao, R. 2017.
\newblock Neural Machine Translation (seq2seq) Tutorial.
\newblock \emph{https://github.com/tensorflow/nmt} .

\bibitem[{Mikolov et~al.(2013)Mikolov, Sutskever, Chen, Corrado, and
  Dean}]{mikolov2013distributed}
Mikolov, T.; Sutskever, I.; Chen, K.; Corrado, G.~S.; and Dean, J. 2013.
\newblock Distributed representations of words and phrases and their
  compositionality.
\newblock In \emph{Advances in neural information processing systems},
  3111--3119.

\bibitem[{Myers et~al.(2018)Myers, Furqan, Nebolsky, Caro, and
  Zhu}]{Myers:2018:PUO:3173574.3173580}
Myers, C.; Furqan, A.; Nebolsky, J.; Caro, K.; and Zhu, J. 2018.
\newblock Patterns for How Users Overcome Obstacles in Voice User Interfaces.
\newblock In \emph{Proceedings of the 2018 CHI Conference on Human Factors in
  Computing Systems}, CHI '18, 6:1--6:7. New York, NY, USA: ACM.
\newblock ISBN 978-1-4503-5620-6.
\newblock \doi{10.1145/3173574.3173580}.
\newblock \urlprefix\url{http://doi.acm.org/10.1145/3173574.3173580}.

\bibitem[{Park and Kim(2019)}]{g2pE2019}
Park, K.; and Kim, J. 2019.
\newblock g2pE.
\newblock \url{https://github.com/Kyubyong/g2p}.

\bibitem[{Peissner, Doebler, and Metze(2011)}]{peissner2011can}
Peissner, M.; Doebler, V.; and Metze, F. 2011.
\newblock Can voice interaction help reducing the level of distraction and
  prevent accidents.
\newblock In \emph{Meta-Study Driver Distraction Voice Interaction}, 24.
  Carnegie Mellon Univ.

\bibitem[{Porcheron et~al.(2018)Porcheron, Fischer, Reeves, and
  Sharples}]{Porcheron:2018:VIE:3173574.3174214}
Porcheron, M.; Fischer, J.~E.; Reeves, S.; and Sharples, S. 2018.
\newblock Voice Interfaces in Everyday Life.
\newblock In \emph{Proceedings of the 2018 CHI Conference on Human Factors in
  Computing Systems}, CHI '18, 640:1--640:12. New York, NY, USA: ACM.
\newblock ISBN 978-1-4503-5620-6.
\newblock \doi{10.1145/3173574.3174214}.
\newblock \urlprefix\url{http://doi.acm.org/10.1145/3173574.3174214}.

\bibitem[{Reitmaier et~al.(2020)Reitmaier, Robinson, Pearson,
  Kalarikalayil~Raju, and Jones}]{10.1145/3313831.3376310}
Reitmaier, T.; Robinson, S.; Pearson, J.; Kalarikalayil~Raju, D.; and Jones, M.
  2020.
\newblock An Honest Conversation: Transparently Combining Machine and Human
  Speech Assistance in Public Spaces.
\newblock In \emph{Proceedings of the 2020 CHI Conference on Human Factors in
  Computing Systems}, CHI ’20, 1–12. New York, NY, USA: Association for
  Computing Machinery.
\newblock ISBN 9781450367080.
\newblock \doi{10.1145/3313831.3376310}.
\newblock \urlprefix\url{https://doi.org/10.1145/3313831.3376310}.

\bibitem[{Sato et~al.(2011)Sato, Zhu, Kobayashi, Takagi, and
  Asakawa}]{Sato:2011:SAV:1978942.1979353}
Sato, D.; Zhu, S.; Kobayashi, M.; Takagi, H.; and Asakawa, C. 2011.
\newblock Sasayaki: Augmented Voice Web Browsing Experience.
\newblock In \emph{Proceedings of the SIGCHI Conference on Human Factors in
  Computing Systems}, CHI '11, 2769--2778. New York, NY, USA: ACM.
\newblock ISBN 978-1-4503-0228-9.
\newblock \doi{10.1145/1978942.1979353}.
\newblock \urlprefix\url{http://doi.acm.org/10.1145/1978942.1979353}.

\bibitem[{Sayago, Neves, and Cowan(2019)}]{10.1145/3342775.3342803}
Sayago, S.; Neves, B.~B.; and Cowan, B.~R. 2019.
\newblock Voice Assistants and Older People: Some Open Issues.
\newblock In \emph{Proceedings of the 1st International Conference on
  Conversational User Interfaces}, CUI ’19. New York, NY, USA: Association
  for Computing Machinery.
\newblock ISBN 9781450371872.
\newblock \doi{10.1145/3342775.3342803}.
\newblock \urlprefix\url{https://doi.org/10.1145/3342775.3342803}.

\bibitem[{Sutskever, Vinyals, and Le(2014)}]{sutskever2014sequence}
Sutskever, I.; Vinyals, O.; and Le, Q.~V. 2014.
\newblock Sequence to sequence learning with neural networks.
\newblock In \emph{Advances in neural information processing systems},
  3104--3112.

\bibitem[{Swarup et~al.(2019)Swarup, Maas, Garimella, Mallidi, and
  Hoffmeister}]{swarup2019improving}
Swarup, P.; Maas, R.; Garimella, S.; Mallidi, S.~H.; and Hoffmeister, B. 2019.
\newblock Improving ASR confidence scores for Alexa using acoustic and
  hypothesis embeddings.
\newblock \emph{Proc. Interspeech 2019} 2175--2179.

\bibitem[{Wiggers(2019)}]{wiggers_2019}
Wiggers, K. 2019.
\newblock The Alexa Skills Store now has more than 100,000 voice apps.
\newblock
  \urlprefix\url{https://venturebeat.com/2019/09/25/the-alexa-skills-store-now-has-more-than-100000-voice-apps}.

\bibitem[{Yilmazyildiz et~al.(2016)Yilmazyildiz, Read, Belpeame, and
  Verhelst}]{doi:10.1080/10447318.2015.1093856}
Yilmazyildiz, S.; Read, R.; Belpeame, T.; and Verhelst, W. 2016.
\newblock Review of Semantic-Free Utterances in Social Human–Robot
  Interaction.
\newblock \emph{International Journal of Human–Computer Interaction} 32(1):
  63--85.
\newblock \doi{10.1080/10447318.2015.1093856}.
\newblock \urlprefix\url{https://doi.org/10.1080/10447318.2015.1093856}.

\end{thebibliography}
